\documentclass{acm_proc_article-sp}
\usepackage{graphics,url}
\usepackage{epsfig, subfigure}
\usepackage{authblk}
\usepackage{amsmath}

\begin{document}
\title{TrackMeNot: Enhancing the privacy of Web Search}
\author{Vincent Toubiana}
	\affil{Alcatel-Lucent Bell Labs\\Application Domain\\vincent.toubiana@alcatel-lucent.com} 
	\affil{Work done while a Postdoctoral Researcher at NYU.}
\author{ Lakshminarayanan Subramanian}
	\affil{New York University\\CS Department\\lakshmi@cs.nyu.edu}
\author{Helen Nissenbaum}
	\affil{New York University\\Media, Culture, and Communication\\helen.nissenbaum@nyu.edu}
\maketitle
\begin{abstract}
\label{abs}
 Most search engines can potentially infer the preferences and interests of a user based on her history of search queries. While search engines can use these inferences for a variety of tasks, including targeted advertisements, such tasks do impose an serious threat to user privacy. In 2006, after AOL disclosed the search queries of 650,000 users, TrackMeNot was released as a simple browser extension that sought to hide user search preferences in a cloud of queries. The first versions of TrackMeNot, though used extensively in the past three years, was fairly simplistic in design and did not provide any strong privacy guarantees. In this paper, we present the new design and implementation of TrackMeNot, which address many of the limitations of the first release. TrackMeNot addresses two basic problems. First, using a model for characterizing search queries, TrackMeNot provides a mechanism for obfuscating the search preferences of a user from a search engine. Second, TrackMeNot prevents the leakage of information revealing the use of obfuscation to a search engine via several potential side channels in existing browsers such as clicks, cookies etc. Finally, we show that TrackMeNot cannot be detected by current search bot detection mechanisms and demonstrate the effectiveness of TrackMeNot in obfuscating user interests by testing its efficiency on a major search engine. 
\end{abstract}
\section{Introduction}
Protecting user privacy while providing personalized services is
challenging: information gathered to personalize a service could be
misused and, thus, harm user privacy. Therefore, the benefit of
personalization should be weighed against the risk of sensitive
information disclosure.\\
Web search is a good illustration of the complexity of providing a
personalized service in a privacy-preserving way. Because many search
queries are too ambiguous to accurately express user interest,
personalized query reformulation is often used to reduce the searched
space. In web search, information used to reformulate queries is
mostly based on previous user searches. Consequently, to provide
personalized searches and to identify search patterns, web search
engines keep search logs. These logs pose a real threat to user
privacy since they have the potential to disclose users' identities
along with their potentially sensitive queries.\\
Because search engines are now utilized to retrieve user-related
information, every query could reveal a fair bit of information about
its issuer. The 2006 AOL log disclosure illustrated how search logs --
even those containing no explicit personal identifiers -- could be
analyzed to reveal user identities. Search engines attempted to
address this challenge by anonymizing logs through different processes
\cite{1409222}. Since the search engine must balance data utility and
user privacy, none of these solutions provide effective privacy
protection. An effective anonymization process should destroy every
link between a user's queries, thereby ruining any personalization
attempt.\\
There have been three broad classes of techniques that have been
proposed to enhance the privacy of web search: (a) using anonymization
networks such as ToR~\cite{1314351}; (b) leveraging private
information retrieval techniques~\cite{1576903,
 _Providing_Privacy_through_Plausibly_Den}; (c) obfuscating search
queries~\cite{trackmenot,TMN1}. Each of these classes of techniques
have their own strengths and limitations as we discuss in detail in
Section 2. While anonymization networks can hide the user's IP
address, the user should take caution in preventing personally
identifiable information to leak through several side channels enabled
in Web pages and browsers. In a recent study, 46\% of ToR users
could be identified with their Google accounts~\cite{histogpah}.  A
single web page is associated with several independent streams which
further increases the complexity and delay incurred in using
anonymization network based solutions. One of the basic challenges
with private information retrieval and search obfuscation is the need
to generate fake search queries which look indistinguishable from
genuine search queries. This requires knowledge about several user
query parameters including topics of interest, timing and frequency,
without which it becomes difficult to provide reasonable privacy
guarantees. In addition, both these class of mechanisms do not handle
side-channel attacks where user information leaks through several
channels including clicks, advertisements, cookies.\\
This paper presents the design of TrackMeNot, a system that
enhances the privacy of web search using search obfuscation.
TrackMeNot addresses many of many limitations of the first released version
~\cite{TMN1} which was primarily designed as a simple-to-use search
obfuscation Firefox extension tool. TrackMeNot aims to address two basic
problems: \\
(a) achieve `query indistinguishability' where a search engine cannot
distinguish between user queries and obfuscated queries.\\
(b) prevent adversaries from using side channels and fingerprints to identify TMN queries.\\
 While it does not provide an anonymity-style guarantee, TrackMeNot
 (TMN) aims to provide \emph{Reasonable Doubt} guarantee that
 inhibits search engines to infer the true user search interests. TMN
 leverages history to mimic user timings for propagating obfuscated
 queries and thereby is not vulnerable to timing analysis. By
 preventing side channel leaks, an obfuscator can not be detected by
 its fingerprint, an adversary can not estimate the number of user who
 installed it. As a result, the \emph{Reasonable Doubt} protection is
 extended to all search engine users, even if they do not obfuscate
 their respective web search histories.\\ TMN is implemented as a
 simple browser extension for Firefox and Google Chrome that is
 extremely simple to use and install.  From a usability perspective,
 the biggest advantage of TrackMeNot is that it retains the
 conventional browsing behavior and requires no additional
 infrastructure as required by anonymization techniques.  Users using
 TrackMeNot, will observe no noticeable change in their web
 browsing experience; an important reason why the first release was very
 widely used despite its lack of strong guarantees. We have evaluated
 TrackMeNot on real search engines and show that it can both
 enhance search privacy as well as not be detected by existing search
 bot detectors. TrackMeNot can achieve this guarantee while
 maintaining a low query overhead.\\ The rest of this paper is
 organized as follows: Section 2 offers an overview of the mechanisms
 that enforce search privacy. Section 3 presents search obfuscation
 and explains TMN privacy objectives. Section 4 describes the privacy
 guarantees that obfuscation introduces. Section 5 shows how TMN has
 been designed to cope with side channel leaks. Section 6 presents
 TMN implementation and Section 7 presents the evaluation results of
 TMN.
\section{Background and Related Work}
In this section, we discuss the strengths and limitations of three
main approaches for enforcing web search privacy: anonymity, private
information retrieval and search obfuscation.
\subsection{Anonymity}
Anonymizing search queries is the most common approach for protecting
web search privacy. To provide anonymity, it is first necessary to
delete all pieces of information that could lead to the identification
of the issuer. However, anonymity is not enough to protect privacy, it
is also critical to remove every piece of information that could
eventually link user queries.\\
While identifiers like the IP address
and browser cookies obviously carry identifying information, there are
other pieces of information that could be used to identify users. EFF
recently revealed that the \textbf{User Agent} header and other
browser specific information could be used as a cookie alternative for
tracking browsers~\cite{useragent}. Because every query leaks
information about its issuer as long as queries belonging to a user
can be linked, the risk of user identification exists. The disclosure
of AOL search logs illustrated that search history contains enough
information to identify a user. It is critical to remove every piece
of information that could link queries made by a user.\\
TOR \cite{tor} is a multi-hop network that is used to access the
Internet anonymously. Because traffic is relayed by several nodes
before being decrypted and forwarded to the web, it should not be
possible for a node to link the unencrypted traffic data with the node
that sent it to the network.  TOR can be used to access any service on
the Internet. Since TOR hides only a user's IP address, the user
should cautiously remove the other identifiers that are created and
used by the service she is accessing to. During a recent experiment on
TOR \cite{histogpah}, 46\% of the TOR exit node users were connected
to a Google Account. If a user issues all her searches using the same
Google account, a TOR exit node could easily connect several of the
user's search sessions. Since the number of exit nodes is limited, it
is not unlikely for an exit node to be used several times by the same
user. Therefore, user identity could be compromised through an
inappropriate usage of TOR to search the web.\\
Although users could pay attention to the identifiers that are set by
the service they use, a more realistic approach consist of configuring
the client to automatically delete these identifiers. PWS
\cite{1314351} is a Firefox extension that redirects only web searches
through the TOR network. PWS removes all unnecessary header from the
HTTP headers (including Cookies and User Agent) and filters search
result page active content to prevent search engines from using web
bugs to track users.\\
Complete anonymity cannot be assured unless active content is removed
from the search page \cite{1314351}. In terms of usability, features
that require JavaScript or Flash support (e.g query suggestion, or
video) are systematically disabled. The privacy protection is not
complete since "click-on" search results are not anonymized. As soon
as the user follows a link, her traffic is no longer directed through
TOR. A search engine could redirect clicks to another domain, where a
user would not hide its IP address. Actually, some search engines
already redirect clicks (on their own domain) to record them in user
profiles. Even if the click through rate is low, queries belonging to
the same search session might be linked (for instance, when the same
keyword appears in consecutive queries). Consequently, a click on a
search result might reveal the IP address that is behind several other
queries.  The TOR network \cite{tor} effectively hides a user's IP,
but exit nodes are the Achilles's Heel of onion routing based
networks. Because they could take advantage of their position to
reorder search results, generate click frauds, or exploit browser
vulnerabilities. Only few TOR nodes accept to be used as exit nodes
since they might be held as responsible for the traffic they forward
on the Internet. Some exit nodes are already known and blacklisted by
search engines and can no longer be used to route search traffic. Most
exit nodes are used by a large number of users and generate a
significant number of searches. Therefore, TOR users often receive
CAPTCHA to verify that they are not bot generating
traffic. Furthermore, routing search queries through the anonymous
TOR network significantly increases the delay from 0.5 to about 10
seconds on average \cite{1576903}. This high latency is caused by the
TOR network to prevent time analysis.\\
To get rid of these additional delays, the Firefox extension
GoogleSharing redirects user search traffic through a single hop
before forwarding it to Google. The GoogleSharing network is composed
of proxies maintained by volunteers. Because GoogleSharing now supports 
SSL, proxies do not have to be trusted. GoogleSharing proxies maintain a
pool of Google Cookie IDs that are sent with search queries.  Since
the previous searches made from these IDs are taken into consideration
to rank search results, it is very likely that the user will receive
personalized results not matching her own, but, instead those of other
proxy users. 
\subsection{Private Information Retrieval}
A Private Information Retrieval protocol queries the \emph{i}-th tuple
of a database without clearly disclosing \emph{i}. Private Information
Retrieval protocols to protect search privacy have been recently
proposed in \cite{1576903, _Providing_Privacy_through_Plausibly_Den}.
To hide a search query $q_{i}$, these protocols first compute a set of
queries {$q_{1}$,$q_{2}$,...,$q_{n}$} and then submit to the search
engine: \emph{Q}=$q_{1}$ OR $q_{2}$ ... OR $q_{i}$ OR ... OR
$q_{n}$. Search engine results that are not related to $q_{i}$ are
discarded on the client side, and the filtered search result page is
`then displayed.\\ The challenge of implementing a PIR protocol is to
generate keywords that are as plausible as the user keywords.  If
$q_{1}$,$q_{2}$,...,$q_{n}$ have similar frequencies than $q_{i}$, a
search engine could not guess which of the \emph{k} queries is the
user's\cite{1576903}. However, the average query frequency is not
relevant when the search engine can rely on the past search history to
discard queries that are not consistent with the previous search
behavior.  This issue is partially addressed in
\cite{_Providing_Privacy_through_Plausibly_Den} which calls for
generating queries related to different topics in order to prevent
user profiling. The user query $q_{i}$ is also replaced by a query
that is likely to return the same results, so it cannot be identified
by the queried engine. \\ PIR protocol adapted to web search assumes
pre-computation of plausible queries. This computation requires
significant knowledge of the queried database. Misspelled queries are
particularly complex to hide since the researched term should be
fixed.  Furthermore, the particular pattern of PIR (containing many
``OR'') reveals the PIR protocol user base to search engines.
\subsection{Obfuscation}
Obfuscation injects artificially-generated data into the set of data to protect. Unlike solutions mentioned above, obfuscation cannot protect a particular data but, instead, can reduce the probability that each data is made by a user. It is critical for the obfuscator to generate data that cannot be distinguished from user data. For instance, to hide the driving habits of drivers, realistic data are generated from traces provided by GPS driver records \cite{1560010} or trip planners \cite{1655204}.
Authors of \cite{1655204} made a qualitative evaluation of their approach and obtained good results against human adversaries.
TrackMeNot \cite{trackmenot,TMN1} is a browser extension that enforces privacy through obfuscation: TMN frequently queries a search engine to mask real user's queries.
\subsubsection{TrackMeNot initial design flaw}
The very first release of TMN has received criticisms \cite{schneier}.
Although all these criticisms have been addressed in the following month, we report them to emphasize the progress that has been made since the release of the first version.\\
\emph{Number of words in the Dictionary:} The first released version of TMN used a static list of keywords to generate search queries. Therefore, a search engine knew that any search that was not in this dictionary had to be user's search.  \\
\emph{Keywords specialization: }To cover a large set of possible queries, TMN used a large set of keywords that were not topic-focused. TMN obfuscation was similar to random noise and could be filtered without losing much information about user profiles.\\
\emph{Bandwidth consumption:} In 2006, bandwidth was a more critical resource than it is now. TMN bandwidth consumption was relatively high and could degrade the user's browsing experience. Although TMN bandwidth consumption has slightly increased, the impact is now negligible compared to common web browsing activities. Loading a simple web page generates many queries to ad content networks, some of them (like videos or flash animations) consume more bandwidth than a TMN query.
\subsubsection{Attack against TMN}
In \cite{TMNAttack} authors use Machine Learning techniques to filter TrackMeNot searches. Although the filter performs a very well and identifies TMN search queries with very high precision, it also classifies, on average, half of user queries as made by TMN. It is worth noting that in all of the instances  in which TMN was attacked the same default settings were used and browsers were never restarted during the attack, thereby preventing TMN from updating the keywords list through the RSS feed. Since this attack, TMN has been updated to include more feeds, as well as randomly click-on query suggestion. This type of attack would, therefore, be even less effective against the new version of TMN.
\subsection{PIR and Obfuscation comparison}
PIR protocols and the obfuscation scheme seem similar as they send
obfuscating queries to protect user's privacy. Beyond this similarity,
these approaches have different objectives and constraints. We
emphasized these differences in the particular context of web search
privacy.\\
To hide the query $q_{i}$, PIR protocols submit a composed
query \emph{Q}=$q_{1}$ OR $q_{2}$ ... OR $q_{i}$ OR ... OR $q_{n}$.
Instead of sending one long query, an obfuscator sends sequentially
$q_{1}$, then $q_{2}$,.. then $q_{i}$ ... and $q_{n}$.  What seems to
be a technical detail has in fact fundamental consequences: a PIR
protocol only protects the search privacy of people using it whereas
an obfuscator protects the privacy of every search engine user; even
those that do not use it.  With a PIR protocol, a user cannot pretend
that she did not issue $q_{i}$ unless she issued ``$q_{1}$ OR $q_{2}$
... OR $q_{i}$ OR ... OR $q_{n}$''. On the other hand, she could argue
that $q_{i}$ is an obfuscating query even if she never used an
obfuscator.\\
PIR protocols prevent a search engine from filtering
keywords through a time analysis. Nevertheless, time analysis could
still reveal a bit of information that could be used to profile
users. For instance, recognizing search timing pattern in a large
community of users could let the search engine to determine the events
a user is attending when she is not making her regular searches. Such
information could reveal critical details such as user religion or
political opinion.  Furthermore, with obfuscation, incertitude can be
total; it is plausible that all user queries have been generated by
the obfuscator, but it is also possible that none of them have. On the
other hand, with PIR, the adversary knows
that 1 out of n queries is the user's.\\
In the next section, we leverage this incertitude to protect search
privacy. Although one may fear that a large adoption of TMN would
cause a flood of queries on search engines, Section 4 demonstrates
that a popular obfuscator could provide Privacy beyond a Reasonable
Doubt by sending a negligible number of queries.

\section{TrackMeNot Search Obfuscation}
The basic design philosophy of TrackMeNot is to achieve two
objectives:\\ 
{\em Query Indistinguishability:} A search engine should
be unable to distinguish genuine user queries and obfuscated
queries.\\ 
{\em Side channel leakage prevention:} Prevent systematic
identification of TMN queries based on fingerprints.\\
In this section we only focus on query indistinguishability and deal
with side channel leakage in Section 5.  In the absence of any side
channel leaks, a user from the perspective of a search engine can be
characterized only using a string of search queries. The basic goal of
search obfuscation is to achieve query indistinguishability to make it
difficult for a search engine to distinguish between genuine and
obfuscated queries. In other words, search obfuscation makes the
analysis of one's search history hard enough to dissuade any community
like AOL Stalker~\cite{aolstalker} from identifying query issuers.\\
We assume that the search engine may use three basic forms of data
analysis to potentially distinguish user requests from obfuscated
requests: (a) topic based analysis; (b) timing based analysis; (c)
frequency based analysis. In topic based analysis, a search engine can
categorize search requests across different topics and use data mining
analysis across history to identify user specific topics.  Timing
based analysis leverage timing information of user requests and
frequency based analysis leverages frequency of requests across topics
and query popularity on the Web.\\
Query indistinguishability is fundamentally different from traditional
anonymity models in that while anonymization tries to destroy every
link between the user and her search queries, obfuscation makes these
links less valuable by diluting them. Even in the face of query
indistinguishability, a search engine can learn a basic set of
features about the user such as: login name, IP address. From a privacy 
perspective, we are only aiming to hide the user queries and specific
user interests but not all the user's credentials.\\
Search engines publish a broad category of fairly general topics such
as ``Entertainment'', ``Travel'' etc. across which they classify
information sources. We assume that there exists a universe $U$ of general
topics that is publicly available; in fact, our implementation leverages
one such universe of 600 topics from the Yahoo profiles. TrackMeNot
can be tailored to support one of two types of Query indistinguishability
models:\\
{\em Topic-exposed Query indistinguishability:} A user search queries span
$m$ broad topics $T_1, \ldots T_m$. In topic-exposed query indistinguishability,
the search engine can easily infer the $m$ broad topics of interest of
a user but will be unable to distinguish genuine and obfuscated queries
within these topics. In this case, the obfuscated queries are generated
across the same broad topics as the original topics.\\
{\em Topic-obfuscated Query indistinguishability:} In this model, we
aim to both provide query indistinguishability as well as hide the
user search interests across a larger set of $n$ topics where $n>m$.\\
We make two important observations. First, the topic-exposed model is
a realistic model since most of the published search-engine profiles
are across very broad topics where revealing a user interest on a
broad topic may not convey much to a search engine about specific user
interests within the topic. Second, topic-exposed query
indistinguishability is much easier to achieve which a much smaller
set of queries in comparison to topic-obfuscated query
indistinguishability. To achieve topic-obfuscated query
indistinguishability in the face of frequency, timing and topic
analysis incurs a much higher query overhead.
\subsection{Topic-exposed Query Indistinguishability}
In this model, TrackMeNot aims to protect against frequency and
timing based analysis of search queries.  The solution approach of TMN
is to consider a universe $U$ of topics as published by search engines
such as Yahoo! which classify user profile across a specified set of
topics.  Given $U$, TMN learns the genuine user profile of topics, and
issues obfuscated queries within the same topics.  To identify
candidate queries, TMN uses a large online repository of RSS
feeds. TMN pre-classifies and pre-identifies the topic of each RSS
feed across the universe $U$. TMN identifies a set of RSS feeds corresponding
to the user search interests and allows the user to pick her favorite
RSS feeds within the set. The next step is to identify n-grams from the
chosen RSS feeds as candidate obfuscation queries. \\
{\bf Frequency analysis:} To thwart basic forms of frequency analysis
from a search engine, TMN maintains a frequency profile (as a
cumulative distribution) across three granularities: frequency of
queries across topics, frequency of keywords within topics and
relative popularity of query n-grams.  Across these three dimensions,
TMN maintains an obfuscation profile which has a similar
relative frequency profile across topics and across keywords within
topics.  To obfuscate specific keywords which may be used repetitively
across queries, TMN leverages a set of obfuscated keywords and follows
a similar frequency pattern as the original keywords. TMN could also
measure the relative popularity of the $n-$gram using published
datasets such as the Linguistic Data Consortium (LDC) and generate
obfuscated queries which have similar relative frequencies.
To model the case where user interests may change with time, TMN can
maintain the user topic profile as a sliding window.\\
{\bf Timing analysis:} TMN uses user's history to be resilient to
timing analysis. The frequency of TMN obfuscation queries is
comparable if not less than the frequency of genuine user queries. TMN
maintains a weekly and daily profiles of timing information of user
requests and attempts to approximately replicate both the timing and
the inter-arrival times across genuine user requests to issue
obfuscated requests; exact replication does expose TMN to timing
analysis -- hence we add a certain amount of randomness to the
inter-arrival times.  In addition, TMN maintains active and inactive
periods of users and issues queries only during active periods; TMN is
activated only when the browser is active. To handle the temporal
correlation across queries, TMN does not intersperse obfuscated
queries with a temporally correlated stream of queries from a user on
a single topic. Using the distribution of temporal correction across
real queries, TMN can correspondingly issue a temporally correlated
stream of obfuscated queries across target topics.
\subsection{Topic obfuscation}
In topic-obfuscated query indistinguishability, TMN needs to both infer
the user topic interests as well as construct a list of obfuscation topics
to hide the user topics.
Given the universe $U$ of topics as published by a search engine,
consider a user whose search requests span $m$ topics. We use an
approach similar to \cite{_Providing_Privacy_through_Plausibly_Den} to
identify a target list of $n$ topics for each user across which we
anonymize the genuine $m$ topics.  
To generate obfuscated queries on
target topics, TMN uses the user-defined list of RSS feeds across
the target topics. Most RSS feeds are topic focused and one can find
RSS about a given topic very easily using a RSS search engine such as
Google Reader or use a Twitter search about a particular topic.
Because RSS feeds are dynamic and more reflective of real-time trends,
we extract keywords from these feeds to construct queries. We believe
RSS-feeds based TMN queries are more likely to follow search trends
than using any static corpus. To identify keywords in an RSS feed, TMN
focuses on RSS titles and then extracts words starting with or
containing a capital letter.  Capital letter starting words are either
names or words which have been emphasized by the web content writers
because they are highly informative about the article
content. Therefore, instead of using complex artificial intelligence
and semantical analysis, TMN leverages content writers information
analysis to distinguish valuable keywords.  TMN allows users to choose
the $n$ target topics and the corresponding RSS feeds to obfuscate her
queries; this allows user-customization in the search obfuscation
process.  \\
To handle frequency analysis in this model, TMN should choose a list
of target obfuscated topics that emulate both long term interest
topics as well as time-varying interests.  If analysis of user search
behavior reflects that some interests are expressed at a constant
frequency (e.g: a user checking every week sports results), TMN should
provide obfuscation queries about (at least) one target topic that is
similarly distributed in time. Similarly, different topics may have
slightly varying frequencies of query requests; for each frequency
range, TMN chooses at least one target topic to emulate a similar
frequency pattern. \\
In the topic obfuscation model, TMN needs to choose at least one
obfuscation topic to hide a real user topic and hence the number of
queries may be much higher than the topic exposed model.  In general,
we note that achieving strong privacy guarantees in the topic
obfuscation model in the face of a powerful oracle is fundamentally
hard. If the obfuscation topic and the actual topic have fundamentally
different semantics, providing absolute guarantees across topics is
impossible and one would need a strong computational linguistic model to argue
the strengths and limitations of search obfuscation. What TMN tries to
defend against is only against a simplistic query analysis model which we
believe is reflective of current search engine practices \cite{1718540}.
\begin{table}[ht]
\caption{Notation} 
	\label{label:Table Notation}
\centering 
\begin{tabular}{|c | l|} 
\hline\hline 
Name & Description \\ [0.5ex] 
\hline 
$G_{q_{i}}$  & The event $q_{i}$ is generated \\
$A_{q_{i}}$  & The event $q_{i}$ is flagged as generated \\ 
P($A_{q_{i}}$)  & Probability for $q_{i}$ to be flagged\\& as artificial ($A_{q_{i}}$=1)\\  
$Ob$ & The user is suspected to obfuscate her searches \\
$p_{Ob}$  & Probability that the user is suspected\\& to obfuscate her searches (P(Ob)) \\
$\tilde{X}$ & Estimated number of artificial queries \\
$\tilde{Y}$ & Estimated number of user queries \\
X & Number of artificial queries \\
Y & Number of user queries \\
\hline 
\end{tabular}
\label{table:nonlin} 
\end{table}
\section {Adversary Models and Privacy Guarantees}
In this section, we define three types of attacks that match different adversaries and give an explanation of the mitigation of TMN against these adversaries.
\subsection{Adversary Model}
We consider an obfuscator taking as input \emph{H}=\{$q_{1}$,$q_{2}$,...,$q_{Y}$\} and outputting \emph{OH}=\{$q_{1}$,$q_{2}$,...,$q_{Y}$, $q_{Y+1}$,...$q_{Y+X}$\} .
The obfuscated search history OH contains \emph{Y} real queries and \emph{X} decoys.\\
The adversary aims at de-obfuscating the search history by filtering the \emph{X} decoys. The adversary just knows the numer of submited queries  \emph{X}+\emph{Y} but does not know the values \emph{X} and \emph{Y}. For each query $q_{i}$ in \emph{OH}, the adversary computes P($A_{q_{i}}$). It is important to notice that P($A_{q_{i}}$) is not the probability that $q_{i}$ is artificial but, rather the probability that it is flagged as artificial. Similarly, \emph{P(Ob)} is not the probability that the user is obfuscating his search but, rather, the probability that the obfuscation will be suspected.\\
We consider three types of attacks that differ on their objectives and the resources and knowledge that the attacker has with regard to the victim.
\subsubsection{Attack I based  on a single query}
The attacker considers a specific query in the search history of a user and evaluates it if this query has been issued by TMN or by the user himself. This kind of attack is typically prevented by covering any potential side channel that would let the attacker know with certitude that a query has been made by the user. This attack could be run by an adversary trying to establish the list of users who have issued a specific query.
\subsubsection{Attack II based on a query set}
The set of queries issued by a user account is analyzed, and the queries are filtered based on the probability that they have been issued by the user. The attacker fixes an arbitrary threshold and filters all the queries that it considers generated. In this attack scenario, the attack can try to detect user search sessions to improve filter performances. This attack can be performed by a de-anonymizing attacker trying to find the identity of someone having issued the set of searches.
The Machine Learning based attack described in \cite{TMNAttack} is an implementation of this attack.
\subsubsection{Attack III based on a user profile}
The adversary knows a set of new queries and the user profile. The adversary filters these queries based on the likelihood that they have been issued by the user considering his profile and semantic features. The attacker could then use this filtered set of queries to update the user profile.
\subsection{Generated Query Indistinguishability}
An obfuscator provide query indistinguishability if:
\begin{equation*} \forall q_{i} \in \emph{OH},  P(A_{q_{i}}|Ob \wedge G_{q_{i}}) = P(A_{q_{i}}|Ob \wedge \overline{G_{q_{i}}}) =\frac{\tilde{X}}{(\tilde{X}+\tilde{Y})} \end{equation*}
The first release of TMN did not provide \emph{Generated Query Indistinguishability}. Indeed, search queries that are associated to suggest queries could not have been generated by TMN  and, therefore, a user could not have claimed that they had been issued by the obfuscator.  \\
This occur when no feature can distinguish a real query from a generated one. If the obfuscator has this property, the search history is protected against \emph{Attack I}.\\
To use this with regard to semantic features, one could apply the method proposed by \cite{ _Providing_Privacy_through_Plausibly_Den} to generate keywords that can be connected to targeted categories by the search engine profiling system. In Section 5, the implementation details illustrate how TMN handles side channel attacks. 
\subsection{Reasonable Doubt}
$\epsilon$ is the Reasonable Doubt that cannot be tolerated when accusing a user to issuing a query.
Obfuscation achieves the Reasonable Doubt $\epsilon$ objective if:
\begin{equation} \forall q_{i} \in \emph{H}, P(A_{q_{i}} \wedge  Ob)\ge\epsilon \end{equation}
 If the obfuscator reaches the \emph{Reasonable Doubt}, there should be no query in user search history that she cannot claim as artificial. 
\subsubsection{$\epsilon$-overhead Obfuscation}
We create a condition for an obfuscator providing \emph{ Generated Query Indistinguishability} to provide Reasonable Doubt. Applying \emph{Law of Total Probability }, we have:
\begin{equation} \begin{aligned}
&P(A_{q_{i}} \wedge  Ob) =  P(A_{q_{i}}|Ob) . p_{ob} + P(A_{q_{i}}|\overline{Ob}).\overline{p_{ob}}
\end{aligned}\end{equation}
$P(A_{q_{i}}|\overline{Ob})$ is the probability for query $q_{i}$ to be flagged as generated by the adversary supposing that the latter does not obfuscate her searches. Therefore, $P(A_{q_{i}}|\overline{Ob})$ = 0 and  
\begin{equation}\begin{aligned}
&P(A_{q_{i}} \wedge Ob) =  P(A_{q_{i}}|Ob) . p_{ob} 
\end{aligned} \end{equation}
Assuming \emph{Generated Query Indistinguishability} property:
\begin{equation} \begin{aligned}
&P(A_{q_{i}} \wedge Ob) = \frac{p_{ob}.\tilde{X}}{(\tilde{X}+\tilde{Y})} 
\end{aligned} \end{equation}
Therefore, \emph{Reasonable Doubt} is equivalent to
\begin{equation}\begin{aligned}
   & \forall q_{i} \in \emph{H},  \frac{\tilde{X}}{(\tilde{X}+\tilde{Y})} . p_{ob} \ge \epsilon \\
   & \Leftrightarrow  \tilde{X} \ge \frac{\tilde{Y}.\epsilon}{(p_{ob}-\epsilon)}\\
\end{aligned} \end{equation}
Furthermore, X + Y = $\tilde{X}$  + $\tilde{Y}$\\
\begin{equation} \begin{aligned}
(5)&  \Leftrightarrow  X+Y-\tilde{Y} \ge \frac{\tilde{Y}.\epsilon}{(p_{ob}-\epsilon)}\\
   & \Leftrightarrow  X \ge \tilde{Y} . \frac{p_{ob}}{(p_{ob}-\epsilon)} - Y
\end{aligned} \end{equation}
The number of queries to generated is not proportional to the number of user queries, but rather to the estimated number of user queries.  
\subsubsection{Reasonable Doubt against conservative adversary}
Because an adversary should not be capable of evaluating either \emph{X} or $p_{ob}$ ( properties not provided by PIR since both the use and the number of generated queries are disclosed) an obfuscator can leverage this uncertainty and force adversaries to consider that each query is artificial.
A conservative adversary especially cannot tolerate having a generated query flagged as human generated. Considering the example of the crowd de anonymizing search logs, information extracted from an artificial query could accuse the wrong user.\\
If the obfuscator cannot be identified by bogus queries, the adversary ignores who is obfuscating its search history. $p_{ob}$ should be the same for every user and is based on obfuscator popularity. 
The more popular the obfuscation the fewer queries have to be issued because $p_{ob}$ increases. Assuming that obfuscation becomes the norm (for example, if it becomes standard in major browsers), a conservative adversary should assume $\tilde{p_{ob}}$ = 1. \\
An adversary is only capable of observing \emph{X+Y}. To prevent an adversary from estimating \emph{X}, the number of generated queries should be completely random for each user and, but, follow a realistic time distribution. $X_{min}$, especially should not be too high to prevent the detection of user which are not obfuscating their searches (those that generate less than $X_{min}$ queries) and $X_{max}$ should not be too low or the adversary would know that \emph{X+Y-$X_{max}$} queries are real. Under such assumptions, a conservative adversary that cannot measure \emph{X} should estimate: $\tilde{Y}$ = 1 .
Therefore, if the obfuscator is popular enough for the conservative adversary to estimate that every one may utilize it, \emph{Reasonable Doubt} $\epsilon$ is guaranteed even with  \emph{X = 0}.
Consequently, a popular obfuscator providing the \emph{Generated Query Indistinguishability} property would grant a reasonable doubt to every search engine user -- even to those that do not obfuscate their searches.
\subsection{($\alpha$,$\beta$) Resiliency }  
For a filter \emph{F}, the true negative rate is given by 
\begin{equation} \begin{aligned}
 TNR = \frac{Tn}{(Tn+Fn)}
\end{aligned} \end{equation}
TNR is the user query ratio in the complete set of filtered queries.
We say that the search obfuscator is $\alpha$ resilient for 0<$\alpha$<1 if, in order to filter the generated query with a TNR $\geq$$\beta$, the search engine should drop at least $\alpha$ of the real query. \\
It estimates what it would cost the search engine in terms of user queries to remove the noise. Since the search engine can tolerate a certain volume of noise in user search history, it does not have to drop queries unless the amount of noise is beyond what it is acceptable. We denote by $\beta$ this acceptable volume of noise.  Intuitively, $\beta$ should be small, otherwise the search engine could infer false information from the profile. 
The ($\alpha$,$\beta$) Resiliency protects the user against Attack II by forcing the attacker to drop legitimate queries. This property could also mitigate Attack III, though it cannot take into consideration the user's profile. The results obtained in \cite{TMNAttack} illustrates TMN resiliency.

\section{Preventing Side Channel Leaks}
Even if generated queries are semantically similar to user queries, technical features could allow search engines to identify some user queries. In this section, we provide an exhaustive list of technical details and patterns that could leak information about real queries. Then, we explain how these leaks are prevented by TMN.
\subsection{Side Channels}
 Search timing is user-specific and varies according to the hour and the day; TMN queries should have a similar time distribution.\\
 The HTTP header contains information about the browser, user operating environment and preference and the URL of the last visited website. These fields must be carefully set by TMN to correspond to those of regular queries. New features and advanced search interfaces also introduce new side channels that could be used to identify user interactions. To display the search result page, the browser downloads many elements, each of which is obtained through a separate request; these requests should be made also when the query is artificial. \\
Active content embedded in search result pages is used to enhance search interface with features such as search suggestions. If the obfuscator does not handle active content like the user, the search engine could identify obfuscating queries via technical features. Furthermore, these new features are utilized by the user and should be similarly used by TMN. The set of all interactions that a user can have with a search engine and that are recorded should be supported by TMN. Finally if TMN does not generate clicks on search results, a search engine could identify searches followed by clicks as user searches.
 In the rest of this section, we detail how TMN prevents the use of side channels to filter artificial queries.
\subsection{Query scheduling}
TMN is integrated with the browser and does not run as a standalone program. As a result, TMN does not generate queries unless the browser is actually open. `Query Bursts' are triggered whenever a real search is issued, thus making it harder for a search engine to evaluate search patterns.\\
 To improve search scheduling, TMN can analyze the browser search history (by querying the local history) or web search history (by parsing the web search history webpage) and identify search patterns in order to map them. 
\subsection{HTTP Header}
\label{sec:HTTP Header}
Miscomputed Cookie and User Agent headers could be used to flag TMN queries. Since queries are issued from a user's browser, Agent and Cookies are computed by the browser and not by the obfuscator, thereby preventing computation error. \\
\emph{url:} TMN uses Regular Expression to catch the search URL and then use this URL to issue new queries. Therefore, when user queries a local version of a search engine, TMN queries the exactly same URL. \\
\emph{referrer:} If two consecutive TMN queries are issued on the same search engine, the referrer of the second query is the URL of the first query result page. Otherwise, the referrer is set to the URL of the last website `visited' by TMN after a click through. The referrer is never set according to the websites that a user visits. Otherwise, TMN would leak information about user browsing sessions.
\subsection{External element downloading}
TMN  loads the search result page in a collapsed browser element. Every element of the result page is downloaded as if it were displayed in a tab. Fingerprints that TMN queries leave in search engine logs are the same as these user queries. Therefore, even by correlating search and request on content hosting servers, search engines cannot filter artificial queries. 
\subsection{Favicon} 
On a browser tab, a favicon is also displayed along the title of the page. To prevent distinction based on the traffic pattern, the favicon is downloaded by TMN when the page is rendered. This icon is rendered in TMN menu bar and is therefore downloaded as if it was rendered in the location bar: Firefox checks the cache to see if the icon is available before downloading it. 
If a valid entry of the favicon is present in the cache; it is not downloaded. The search engine can not use cache information leakage to flag real user queries either. 
\subsection{Active Content Handling}
 In this new version of TMN, active content is supported on the search result page: the JavaScript embedded in search result pages is downloaded and then executed in the browser. This might be a concern if some malicious JavaScript is executed by TMN. However, such an event is very unlikely since the executed JavaScript code comes from the search engine webpage.\\
The JavaScript environment in which TMN search results are interpreted should be the same as the environment used to interpret user searches. For instance, when a websearch on Bing is made through an iframe, the javascript embedded in the result page raises an exception when the code checks the length of the DOMWindow history element. The search engine could detect these exceptions and  filter TMN search queries. Therefore, search made by TMN are issued through a browser element which is similar to the browser used by the user to submit her queries.  
\subsection{Query suggestions}
 Instead of mimicking search traffic, TMN mimics interactions with search engine interfaces. This approach is more robust and should be less affected by future change to these interfaces. To implement this, TMN simulates every DOM event that is monitored on the search engine web page. These events are the same as those that are triggered by the user when she is interacting with the search page, thus making it impossible for the search engine to detect queries based on the events. When TMN simulates keystrokes in the search box, query suggestions are automatically requested to the search engine. Therefore, TMN leaves a normal fingerprint of requests for suggestions on the search engine servers. 
\subsection{Click Stream}
 As a real user, TMN sometimes follows search result links and navigates between search results pages by clicking on links. To address security concerns that might be raised if TMN loads untrusted script, TMN never actually downloads content from clicked URLs.\\
Clicks on sponsored links are voluntarily prevented not to interfere with the search engines business model. This could potentially leak information about user searches but it prevents click fraud abuses. The sole solution is for a user not to click on sponsored links related to sensitive searches. 
\section{ Implementation}
TMN  \cite{trackmenot} is implemented as an extension of the popular web browsers Firefox and Google Chrome. It runs as a background process and, thus, does not significantly impact the user browsing experience. Current version of TMN can obfuscate search on AOL, Baidu, Bing, Google and Yahoo!.  \\
Since the very first released version, transparency and ease of use were the values embedded in TMN design. 
\subsection{Values in Design}
TMN is more transparent and lets users see how it acts in background. By clicking on the \emph{Show Frame} button, the user will be presented the browser that submit queries to search engines. Therefore the user can see the queries as they are submitted, the result page that is returned by the search engine and view search result link get colored as TMN clicks on them. 
The option window still provides user with a complete interface letting them select the search engines that are queried by TMN  and set the query frequency. This frequency is set by default to three queries per hour and per engine. Users can change this frequency, TMN will include a feature adapting this frequency to the actual user use of a search engine. Therefore, search engines will not be able to notice an increase in the query frequency when a user installs TMN. The use of RSS feeds to boostrap the list of keywords that are used by TMN is particularly interesting when users are able to edit the feeds that TMN uses. Therefore, TMN includes a editor to add RSS feeds and organize them in categories. This RSS editor is integrated in Firefox as an RSS manager. As a result, a user can add a feed to the TMN RSS list simply by clicking on the RSS logo that appears in the URL box.  
\subsection{Setting up an obfuscating profile}
In order to set-up an obfuscating profile, users need to select a set of RSS feeds that are connected to the list of interests that they want to simulate through TMN. To find a published RSS feed related to a particular topic, users could search on Google Reader for the topic of their choice and then select one of the results.\\
A second approach calls for using Twitter as an RSS feed. This approach can be followed either by setting up a user account that receives tweets related to the topics or by setting up a search on twitter and outputting the result as a RSS feed.
The user should also configure his RSS feeds in categories to map the categorization of the profiling search engine. Users do not show the same level of interest for every topic. They are likely to issue more queries about their preferred topics than about topics that they do not consider very interesting or that are not frequently updated. TMN should similarly show more interest for certain topics than others by generating more queries about them. Once the user has organized his RSS feeds into categories, he'll be able to assign weight to topics. A topic with a greater weight will be the subject of more TMN queries. If the user does not want to assign weights to topics, TMN will analyze the browser search history to extract the queries per topic distribution and use the same distribution.
\section{Evaluation}
In this section, we first overview two search bot detection mechanisms and then evaluate quantitatively and qualitatively the resilience of TMN against these mechanisms. TMN Topic Level Obfuscation is evaluated by examining profiles established by a search engine. 
\subsection{Search Bot Detection}
\subsubsection{Aggressive search bot detection}
\begin{figure}
	\begin{center}
		\includegraphics[width=0.4\textwidth]{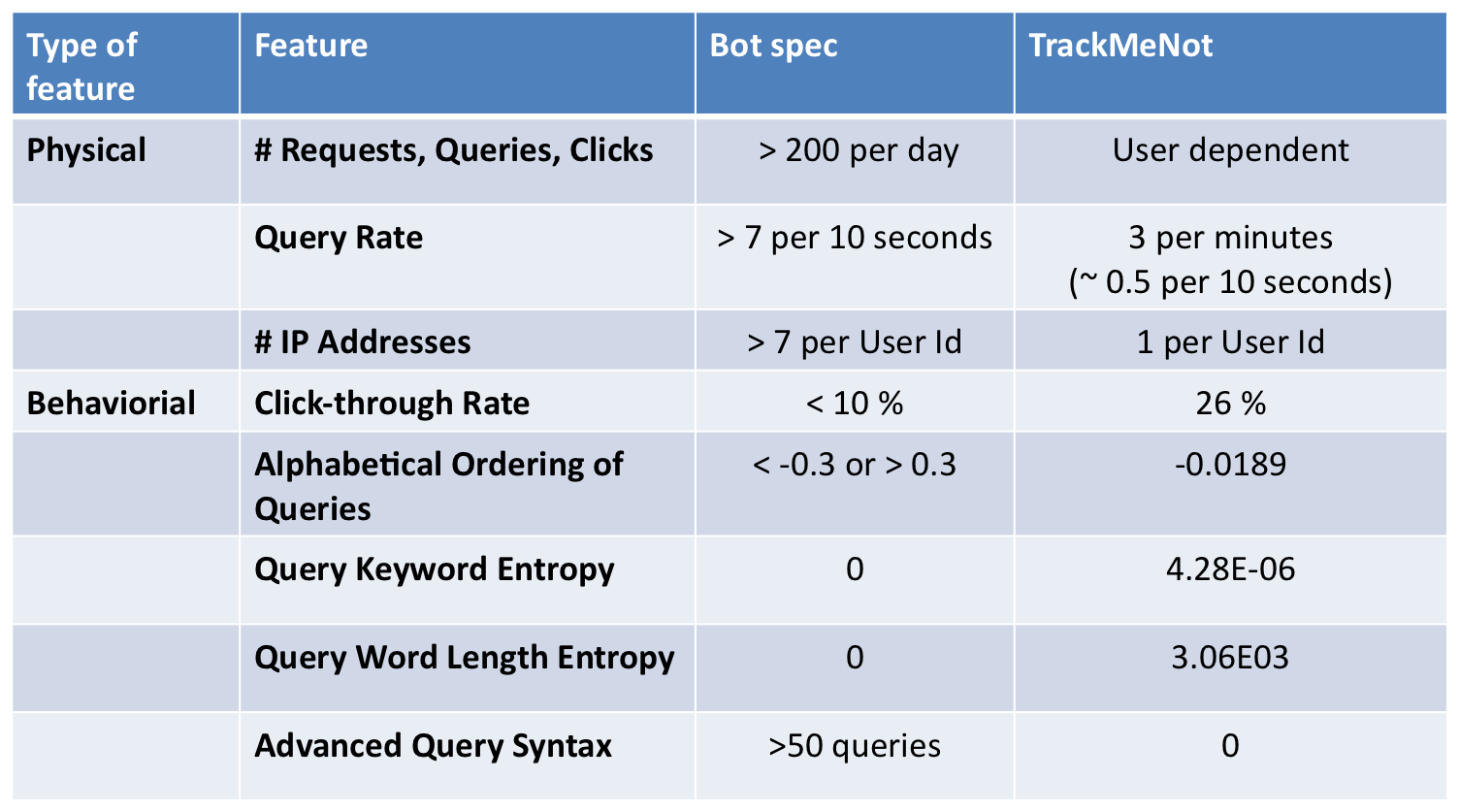}
		\caption{Bot Features}
	\label{fig:Table Airweb 08}
	\end{center}
\end{figure}

\textbf{\emph{Description: }} Buehrer et al\cite{1451985} analyzed search traffic to detect obviously non-human search patterns. They distinguish two sets of features (see Figure \ref{fig:Table Airweb 08}). Physical Model Features reflect a specific traffic patterns while Behavioral Features are related to the content of the search queries. Most of these features focus on a property (keyword length, periodicity, number of keywords...) to evaluate the entropy of user searches. Any behavior which obviously deviates from the norm is flagged as suspicious. \\
\textbf{\emph{TMN Detection: }} We carry out experimentation and estimate the score of TMN for every feature that can be computed. Unfortunately, many features require a large set of human queries to be observed. To compute a score for \emph{Query Word Entropy}, we use the Linguistic Data Consortium (LDC). Computation of Adult Content and Spam scores, require the list of correspondence (word ,weight).\\
From the results reported in Figure \ref{fig:Table Airweb 08}, it appears that TMN can not be detected through any of the feature discovered by \cite{1451985} and that we have been able to observe. Keyword length entropy is estimated based on the exact same formula that is used by \cite{1451985}, though the results are not comparable to those they have obtained. 
\subsubsection{Distributed search bot detection}
\textbf{\emph{Description: }}Features identified by \cite{1451985} detect search bots that expose behavior that is obviously non human. Although these features detect aggressive search bots, they are ineffective against stealth bots that do not expose aberrant behavior.\\
By balancing the load among several members of a BotNet, distributed search bots deceive threshold-based detection mechanisms. But a brutal variation of search traffic might be completely legitimate. For instance, the abrupt interest for Michael Jackson in the few hours following his death, categorized as suspicious by Google, was actually legitimate.\\
SBotMiner\cite{1718540} addresses this challenge in two steps: first it identifies potential search bots by identifying suspicious search activity groups. Then Matrix Based Search Detection removes flash crowds by comparing the entire query/click history of users in suspicious groups. Those that match perfectly are very likely to belong to bots. \\
\textbf{\emph{Search desobufscation:}}Whereas the previous Bot Detection could have been used to identify independently each bot, SBotMiner is used to identify a BotNet. However, SBotMiner mechanism is based on the main assumption that bots are controlled by a remote commander and have similar search activities. Although search activities of two TMN users may show some similarities, TMN search activity should vary very quickly between different users. In fact, if it is appropriately configured TMN should always generate user specific without requiring a learning phase. Such configuration of TMN should not issue the same queries and therefore should not be considered as suspicious a suspicious search activity. Therefore, TMN shall not be identified as a potential BotNet.
\subsection{Experiment}
The previous analyses tend to show that existing solutions to detect search bots shall not flag TMN queries as suspicious. Practically, the use of the obfuscator can not be detected and no features let a search engine distinguish artificial and user queries. From the previous result, it is reasonable to assume that TMN provides query indistinctability. \\The next experimentation shows that TMN also provides Topic Level Obfuscation by evaluating the impact of TMN queries on the profile established by Yahoo!. The search engine Yahoo! has been chosen because it lists the interests that are deduced from user search and browsing history  \cite{yahoointerest}. On this webpage, a user can view the list of interest inferred from his search history but cannot edit this list of interest. To the best of our knowledge, Yahoo! is the sole search engine to provide this list of interest. Other search engines (like Google) do not seem to select displayed ads based on user long-term interests \cite{imc10-guha}.\\
This experiment evaluates the impact of TMN queries on eight Yahoo! profiles. These profiles do not belong to real users and have, therefore, not been used before. In this experiment, each profile is used to send a significant number of queries related to at least one targeted topic. On average, 180 queries are issued from each profile, with a rate of 3 queries per minute. Every hour, the profile is changed and a new set of topics is selected; the IP address of the computer sending the query is also modified. Before presenting the results of the experiment, we describe in details the tools used in the experiment. 
\subsubsection{Experiment Setup}
\textbf{\emph{Topic focused query generator:}}
: To demonstrate topic level obfuscation, we select a list of targeted topics around which interests are expressed. Both Google and Yahoo! have published the list of topics they use to categorize user interests. These lists of topics are broad enough to cover most user interests. Since we observe the impact of TMN on Yahoo! profiles, the Yahoo! topics list is preferred to establish a pool of targeted interests. More precisely, we focus on the 14 root categories plus the 7 `miscellaneous' sub-categories to keep our analysis as broad as possible. Keywords about one of these topics are generated using the tool proposed by  \cite{russell}. 
This tool submits the name of the topic as a query, parses the returned documents, and computes TF and IDF scores of each term occurring in the first thousand retrieved documents. We keep only rare terms that are specific to the targeted topic. Each of  the eight search profiles focuses on two random topics.
It should be noted that this query generation process can be replicated by an adversary to build the list of all obfuscating queries. Therefore this process should normally be coupled with the RSS based query generation process which is dependent of users input.\\
\textbf{\emph{Identity manager:}}
Another Firefox extension has been developed to simplify cookie management. This extension identifies the cookie keys that are used to identify users and stores them. The extension automatically recognizes the name of the current account by parsing the search page and then creates a new entry in the local base. This entry contains the name of the account, cookie values and the randomly selected topics. When an account is selected, the extension replaces the cookie values for the Yahoo! domain with those corresponding to the new account. Then the extension replaces the current list of TMN queries by the list of keywords related to the topic selected for this profile. In order to obtain significant results over a short period of time, we raised the query frequency to three queries per minute.
\subsubsection{Results extraction}
For each profile, we observed the categories and sub categories that are listed. Because Yahoo! simply lists search interest without ordering, it is not possible to know the exact shape of the profile.  We observed that every profile built by established by Yahoo! is different. Because all queries have been issued by TMN, these differences prove that TMN queries are taken into account to build the profiles. Since most categories have sub-categories that are also listed in the list of interests, we approximate the interest that Yahoo! inferred according to the number of subcategories of the targeted topic that appears in the list. For each test, two values are computed: \\
\textbf{\emph{Impact:}} The number of categories and sub-categories listed in the profiles that are related to one of the targeted topics is divided by the overall number of categories related to a targeted topic. The higher the Impact, the more TMN can orient the profile. An Impact of 1 would mean that TMN fully simulates interest for targeted topics. \\
\textbf{\emph{Precision:}} The number of categories and sub-categories listed in the profile that are related to a targeted topics is divided by the total number of categories and sub categories listed in the profile. The higher is the Precision, the more accurate is the topic level obfuscation. If Precision is 1, TMN queries exclusively express an interest for the targeted topics, without affecting any other topic. Result of the experimentation are provided in Table\ref{label:Table Experimentation Result}
\subsubsection{Topic Level Obfuscation Impact}
On average, 40\% of the sub-categories related to a targeted topic is present in the list of interest established by Yahoo!; low results correspond to topics that have only few subcategories (as reported by experiment 2,5,6 and 7). Categories having more specializations are more likely to have better performance in terms of Impact. Thus, the method used to compute the Impact factor might not completely reflect the deduced interests. For instance, the categories ``Government and Military,Romance, Religion and Spirituality, Politics'' have only one sub-category and are therefore less likely to impact user profile. Although we might have expected higher values, these results tend to show that queries generated by TMN impact the categories and sub-categories that are listed in the Yahoo! profile. 
\subsubsection{Topic Level Obfuscation  Precision}
Except for the first experiment, the precision of Topic Level Obfuscation is very low. Several factors explain these results. First, for some topics, there are some mismatches between extracted keywords and the Yahoo! categorization algorithm. For instance, the set of keywords related to the topic ``Military and Government'' ( that was selected in experiments 2) is a list of countries mentioned in document related to International events. But these country names have been considered, most of the time, as travel destinations by Yahoo! profiling system adds many entries to the ``Travel'' category. Secondly, some categories are to broad and systematically listed in profiles. The categories ``Entertainment'',``Small Business and B2B'' and ``Travel'' are listed in test result. These categories are probably too common and could be related to many web searches. Finally, the Yahoo! profiling system has not been evaluated yet. It is therefore hard to know how accurate the algorithm is in identifying user interests. 
\begin{table}[ht]
\caption{Experimentation Results} 
	\label{label:Table Experimentation Result}
\centering 
\begin{tabular}{|c | c c|} 
\hline\hline 
Test & Impact & Precision  \\ [0.5ex] 
\hline \hline 
1 & 0.47 & 0.41 \\ 
2 & 0.09 & 0.03 \\
3 & 0.50 & 0.11 \\
4 & 0.50 & 0.10 \\
5 & 0.57 & 0.12 \\
6 & 0.35 & 0.31 \\ 
7 & 0.43 & 0.08 \\
8 & 0.57 & 0.44 \\
\hline \hline  
Average & 0.44 & 0.20\\[1ex] 
\hline 
\end{tabular}
\label{table:nonlin} 
\end{table}
\section{Conclusion}
Many criticisms have been raised against the use of obfuscation to
protect web search privacy. This paper addresses most of these
criticisms and emphasizes that TMN obfuscation could provide
reasonable doubt protection without overloading search engines. TMN
also addresses several common side channel leaks that occur in web
search.  By experimenting TMN with Yahoo! search profiles, we show
that TMN does impact the inferred search profile and effectively hides
user search interests.  Evaluation with existing search bot detection
mechanism has revealed that TMN search traffic should not be flagged
as automated. We plan to release the new version of TMN for web users
in the near future.

\section{Acknowledgments}
We thank the NYU Privacy Research Group for motivating this research and providing useful feedback to an early presentation of results. We thank Nitesh Saxena, Sai Teja Peddinti, Simson Garfinkel, Antonio Nicolosi and Max Sobell for fruitful and productive conversations. This research was supported by AFOSR MURI award ONR BAA 07-036 and NSF GENI award CNS-0820795.

\bibliographystyle{abbrv}
\bibliography{sigproc-sp}

\end{document}